\documentclass{IOS-Book-Article}

\usepackage{mathptmx}

\usepackage{latexsym}
\usepackage{amssymb}
\usepackage{tabularx}
\usepackage{amsfonts}
\usepackage{listings} 

\usepackage{colortbl}
\usepackage{xcolor}
\usepackage{pst-all} 
\usepackage{pst-poly}
\newpsobject{showgrid}{psgrid}{subgriddiv=1,griddots=10,gridlabels=6pt}
\usepackage{graphicx}
\usepackage{fancybox}
\usepackage{wrapfig}

\usepackage{algorithm}
\usepackage{algpseudocode}

\newcommand{\con}{\wedge} 
\newcommand{\dis}{\vee} 
\newcommand{\alw}{\Box} 
\newcommand{\imp}{\Rightarrow} 
\newcommand{\som}{\Diamond} 

\newtheorem{definition}{Definition}

\newtheorem{corollary}{Corollary}

\newcommand\nonofoot[1]{%
   \begingroup
   \renewcommand\thefootnote{}\footnote{\kern-0.2ex#1}%
   \addtocounter{footnote}{-1}%
   \endgroup
}

\begin{document}
\begin{frontmatter}              

\title{Proposal of a multiagent-based smart environment for the IoT}
\runningtitle{Proposal of a multiagent-based smart environment for the IoT}

\author[A]{\fnms{Rados{\l}aw} \snm{Klimek}%
\thanks{Corresponding Author: Rados{\l}aw Klimek: AGH University of Science and Technology,
al. Mickiewicza 30, 30-059 Krakow, Poland; E-mail: rklimek@agh.edu.pl.}},
\author[A]{\fnms{Leszek} \snm{Kotulski}}

\runningauthor{R.~Klimek and L.~Kotulski}
\address[A]{AGH University of Science and Technology\\ Krakow, Poland}

\begin{abstract}
This work relates to context-awareness of things that belong to IoT networks.
Preferences understood as a priority in selection are considered,
and dynamic preference models for such systems are built.
Preference models are based on formal logic,
and they are built on-the-fly by software agents observing the behavior of users/inhabitants,
and gathering knowledge about preferences expressed in terms of logical specifications.
A 3-level structure of agents has been introduced to support IoT inference.
These agents cooperate with each other basing on the graph representation of the system knowledge.
An example of such a system is presented.
\end{abstract}

\end{frontmatter}

\thispagestyle{empty}
\pagestyle{empty}

\section*{Introduction}
\label{sec:introduction}

\nonofoot{This is a draft/accepted version of the paper:
R.~Klimek, L.~Kotulski: Proposal of a multiagent-based smart environment for the IoT.
\emph{Workshop Proceedings of the 10th International Conference on Intelligent Environments, Shanghai, China, 30th June--1st of July 2014}.
Juan C.\ Augusto and Tongzhen Zhang (Eds.),
ser.\ Ambient Intelligence and Smart Environments,
vol.\ 18,
pp.\ 37--44.
IOS Press.
Available at: \texttt{DOI:10.3233/978-1-61499-411-4-37} or \texttt{http://ebooks.iospress.nl/volume/workshop-proceedings-of-the-10th-international\-conference-on-intelligent-environments}}

The Internet of Things, or IoT, refers to uniquely identifiable objects
able to perform automatic  data transfer over a network and cooperate without any kind of intervention.
Context-awareness is a property related to linking changes in the environment with computer systems,
which are otherwise static.
Important aspects of context are: where you are, who you are with,
and what resources are nearby~\cite{Dey-Abowd-2000}.
Preferences understood as a priority in the selecting  something over
others things are considered.

The contribution of this work is an idea of a smart,
i.e.\ context-aware and pro-active system, which is built using
formal/temporal logic as a method of representing knowledge
and reasoning about inhabitants' behaviors/preferences.
This idea is implemented using a 3-level hierarchy of agents
operating in a graph representation of the IoT.

A car park is considered as an example and defines a graph structure for the IoT.
Dynamic preference models are based on temporal logic,
and they are built on-the-fly by software agents sensing and reacting to users/inhabitants
and preparing user-oriented preference decisions.
Formal logic allows to register behavior in a precise way,
i.e.\ without any ambiguity typically found in natural languages.
``Logic has simple, unambiguous syntax and semantics.
It is thus ideally suited to the task of specifying information systems'' (J.~Chomicki, J.~Saake).
It also allows to perform automatic and trustworthy reasoning,
e.g.\ using semantic tableaux,
to obtain preference decisions for newly observed users/inhabitants.
Logical inference allows to build truth trees,
searching for satisfiability or contradictions.
Agents gather basic information in nodes,
identify users/inhabitants,
observe their behaviors, build logical specifications,
and prepare preference decisions for users/inhabitants.

\section{Context models and preference models}
\label{sec:context-model}

Pervasive computing or ubiquitous computing can be understood
as existing or being everywhere at the same time,
assuming the omni-presence of computing which provides strong
support for users/inhabitants
and makes the technology effectively invisible to the user.
Context-awareness and context modeling is one of the crucial aspects
of pervasive systems
and IoT, which in turn could be understood as a scenario in which objects,
users, inhabitants (or even animals) permanently cooperate.

\begin{figure}[htb]
\begin{center}
\includegraphics[width=.75\textwidth]{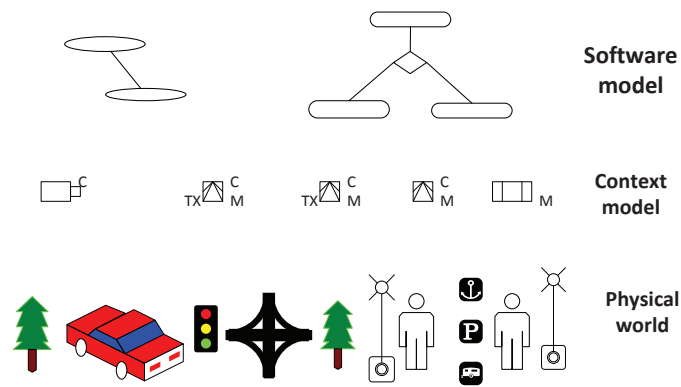}
\end{center}
\vspace{-.8\baselineskip}
\caption{From physical world to context-aware software systems}
\label{fig:context-aware-systems}
\vspace{-.4\baselineskip}
\end{figure}

Context-awareness refers to the interpretation logic that is embedded inside pervasive applications.
This type of computing assumes transfer of contextual information among applications in the IoT network.
The context includes conditions and circumstances that are relevant to the working system.
A sample physical world which creates a context interpreted by
context-aware applications is shown in Fig.~\ref{fig:context-aware-systems},
c.f.\ also~\cite{Bettini-etal-2010}.
The physical world and the context-aware software constitute the smart environment.
The context models are created by different types of sensors
distributed in the whole considered physical area.
Distributed sensors constitute a kind of eyes for software systems.
It follows that smart applications must
both understand the context,
(be context-aware), and be characterized by pro-activity,
meaning they must act in advance to deal with an expected occurrences or situations,
especially negative or difficult ones.
Context-aware systems are able to adapt their operations to
the current context without explicit user intervention.

Preference modeling enables customization of software behavior to users' needs.
The construction of preference models is particularly important in
systems of pervasive computing.
Preference modeling needs formalization and it is discussed in
some works, e.g.~\cite{Ozturk-etal-2005}. The model of preference
might be constructed using fuzzy sets, classical logic and
multi-valued logics. Classical logic, and particularly rule-based
systems, are especially popular.
Non-classical logics, especially temporal logic, are less popular.
On the other hand, temporal logic is a well-established
formalism for describing reactiveness, and meanwhile, the typical
pervasive application should be characterized by reactivity and
flexibility in adapting to changes on the user side.
The variability and change in valuation of logical statements are
difficult to achieve in classical logic,
and that is why temporal logic is proposed.
After building a preference model in formal/temporal logic, one can
analyze it using a deductive approach. The goal is searching for any
contradictions in a model or analyzing its satisfiability.

\emph{Temporal Logic} TL is a branch of symbolic logic and focuses on
statements whose valuations depend on time flows.
It it strongly applicable in the area of software engineering, and
is used for system analysis where behaviors of events are of interest.
TL exists in many varieties; however, considerations in this paper are
limited to the \emph{Propositional Linear Temporal Logic} PLTL,
i.e.\ logic with the linear time structure,
and its semantics could be found in many works,
e.g.~\cite{Wolter-Wooldridge-2011}.
The issue of preference models based on temporal logic are discussed in some works.
For example, in~\cite{Klimek-2013-peccs} some basic notions and definitions
are introduced. The architecture of an inference system is proposed.
The methodology for gathering information about preferences
in the requirements engineering phase is proposed in~\cite{Klimek-2013-icmmi}.
Finally,
in~\cite{Klimek-Wojnicki-Ernst-2013-icaisc},
it is shown that preference modeling could reduce the state space of the agent-based world.

Logic and reasoning are cognitive skills.
Logical reasoning is the process of using
 sound mathematical procedures on given statements to arrive at conclusions.
Although the work is not based on any particular method of reasoning,
the method of semantic tableaux is presented in a more detailed way.
The method of \emph{semantic tableaux}, or \emph{truth tree}
is well-known in classical logic but it can be applied
in modal/temporal logic~\cite{Agostino-etal-1999}.
The method is based on predefined formula decompositions.
At each step of the well-defined procedure,
formulas become simpler as logical connectives are removed.
At the end of the decomposition procedure,
all branches of the received tree are searched for contradictions.
When the branch contains a contradiction, it means that it is \emph{closed}.
When the branch does not contain a contradiction, it means that it is \emph{open}.
When all branches are closed, it means that the tree is closed.
Simple examples of inference trees are shown in Fig.~\ref{fig:truth-trees},
where the adopted decomposition procedure refers to the first-order predicate calculus.

The semantic tableaux method can be treated not only as a method for
system correctness analysis~\cite{Klimek-2012-icaart,Klimek-2012-enase,Klimek-Faber-Kisiel-Dorohinicki-2013-icmmi}
but also as a \emph{decision procedure},
i.e.\ an algorithm that can produce a Yes/No answer as a response to some important questions.
Let $F$ be the examined formula and ${\cal T}$ is a truth tree build for a formula.
\begin{corollary}{}
\label{th:decision-procedures}
The semantic tableaux method gives answers to the following questions related to the satisfiability problem:
\begin{itemize}
\item formula $F$ is not satisfied iff the finished ${\cal T}(F)$ is closed;
\item formula $F$ is satisfiable iff the finished ${\cal T}(F)$ is open;
\item formula $F$ is always  valid iff finished ${\cal T}(\neg F)$ is closed.
\end{itemize}
\end{corollary}
The proof seems relatively easy and it follows from the introduced definitions and rules.

\section{Context-awareness of the IoT}
\label{sec:context-IoT}

The goal of this approach is building preference models on-the-fly,
i.e.\ preference models are created during operation of the system,
and they are the result of users/inhabitants' behavior observations.
Preference models are expressed in terms of temporal logic formulas and
can be dynamically changed during the system operation.

\begin{figure}[htb]
\begin{center}
\begin{pspicture}(11.0,7.0) 
\rput(0.0,0.0){\psframe[linecolor=gray,fillcolor=gray,fillstyle=solid](1.5,1.5)}
\rput(10.1,0.5){\psframe[linecolor=gray,fillcolor=gray,fillstyle=solid](0.9,6.0)}
\rput(1.0,5.0){\psframe[linecolor=gray,fillcolor=gray,fillstyle=solid](5.0,2.0)}
\rput(5.0,0.0){\psframe[linecolor=gray,fillcolor=gray,fillstyle=solid](1.0,3.0)}
\rput(2.0,2.0){\psframe[linecolor=gray,fillcolor=gray,fillstyle=solid](1.5,1.5)}
\rput(7.0,5.0){\psframe[linecolor=gray,fillcolor=gray,fillstyle=solid](2.0,1.0)}
\rput(6.0,0.0){\psframe[linecolor=gray,fillcolor=gray,fillstyle=solid](1.0,0.4)}
\psline(0,0)(3,0)
\psline(4,0)(7,0)
\psline(8,0)(11,0)
\psline(11,0)(11,0.5)
\psline(11,6.5)(11,7)
\psline(6,7)(11,7)
\psline(0,7)(1,7)
\psline(0,5)(0,7)
\psline(0,4)(0,0)

\rput(8.0,0.5){\psframe(1.0,0.45)}
\rput(8.0,1.0){\psframe(1.0,0.45)}
\rput(8.0,1.5){\psframe(1.0,0.45)}
\rput(8.0,2.0){\psframe(1.0,0.45)}
\rput(8.0,2.5){\psframe(1.0,0.45)}

\rput(6.1,0.5){\psframe(0.9,0.45)}
\rput(6.1,1.0){\psframe(0.9,0.45)}
\rput(6.1,1.5){\psframe(0.9,0.45)}
\rput(6.1,2.0){\psframe(0.9,0.45)}
\rput(6.1,2.5){\psframe(0.9,0.45)}
\rput(8.0,3.5){\psframe(1.0,0.4)}
\rput(8.0,4.0){\psframe(1.0,0.4)}
\rput(8.0,4.5){\psframe(1.0,0.4)}\rput(8.5,4.7){\small p018}
\rput(6.1,3.5){\psframe(0.9,0.4)}
\rput(6.1,4.0){\psframe(0.9,0.4)}
\rput(6.1,4.5){\psframe(0.9,0.4)}
\rput(5.0,3.5){\psframe(1.0,0.4)}
\rput(5.0,4.0){\psframe(1.0,0.4)}
\rput(5.0,4.5){\psframe(1.0,0.4)}\rput(5.5,4.7){\small p015}

\rput(9.1,0.5){\psframe(0.9,0.45)}
\rput(9.1,1.0){\psframe(0.9,0.45)}
\rput(9.1,1.5){\psframe(0.9,0.45)}

\rput(4.0,0.5){\psframe(0.9,0.45)}
\rput(4.0,1.0){\psframe(0.9,0.45)}\rput(4.5,2.2){\small p010}
\rput(4.0,1.5){\psframe(0.9,0.45)}
\rput(4.0,2.0){\psframe(0.9,0.45)}
\rput(4.0,2.5){\psframe(0.9,0.45)}

\rput(2.0,0.5){\psframe(1.0,0.45)}
\rput(2.0,1.0){\psframe(1.0,0.45)}
\rput(2.0,1.5){\psframe(1.0,0.45)}

\rput(1.0,2.0){\psframe(0.9,0.45)}
\rput(1.0,2.5){\psframe(0.9,0.45)}
\rput(1.0,3.0){\psframe(0.9,0.45)}

\rput(1.0,3.6){\psframe(0.9,0.45)}
\rput(2.0,3.6){\psframe(0.9,0.45)}

\rput(7.5,0.0){g1}
\rput(3.5,0.0){g2}
\rput(0.0,4.5){g3}
\end{pspicture}
\end{center}
\caption{A sample parking space as a graph structure}
\label{fig:graph-parking-space}
\end{figure}
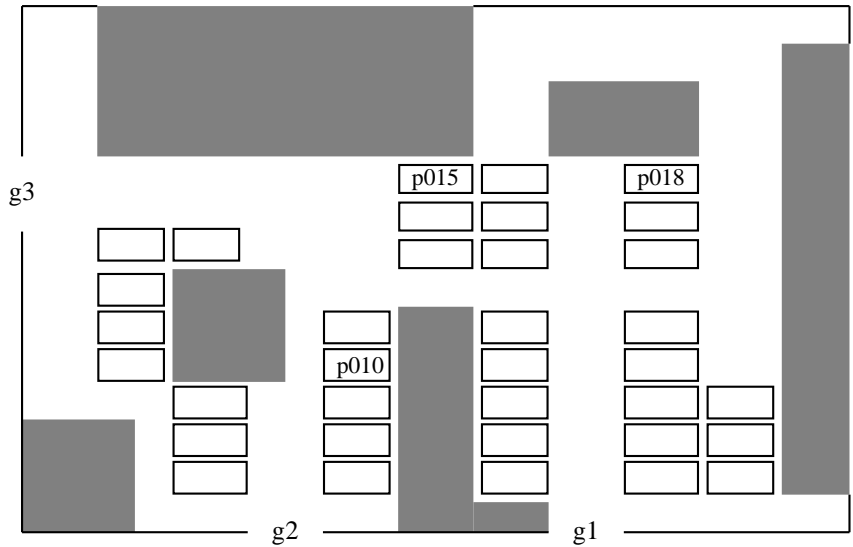
\begin{figure}[htb]
\centering
\begin{pspicture}(6.5,2.5) 
\psset{linecolor=blue,fillcolor=blue,fillstyle=solid}

\psline[linecolor=gray,fillcolor=gray,fillstyle=solid](0,0)(6.4,0)
\rput(3.2,.2){\textsc Graph layer}
\psline[linecolor=gray,fillcolor=gray,fillstyle=solid](0,.4)(6.4,.4)

\rput(0.0,0.5){\psframe(.4,.6)}
\rput(0.5,0.5){\psframe(.4,.6)}
\rput(1.0,0.5){\psframe(.4,.6)}
\rput(1.5,0.5){\psframe(.4,.6)}
\rput(2.0,0.5){\psframe(.4,.6)}
\rput(2.5,0.5){\psframe(.4,.6)}
\rput(3.2,.8){\textsc A1}
\rput(3.5,0.5){\psframe(.4,.6)}
\rput(4.0,0.5){\psframe(.4,.6)}
\rput(4.5,0.5){\psframe(.4,.6)}
\rput(5.0,0.5){\psframe(.4,.6)}
\rput(5.5,0.5){\psframe(.4,.6)}
\rput(6.0,0.5){\psframe(.4,.6)}

\rput(0.8,1.2){\psframe(.6,.6)}
\rput(2.0,1.2){\psframe(.6,.6)}
\rput(3.8,1.2){\psframe(.6,.6)}
\rput(5.0,1.2){\psframe(.6,.6)}
\rput(3.2,1.5){\textsc A2}

\rput(1.8,2.0){\psframe(3,.6)}
\rput(3.2,2.3){\textsc A3}
\end{pspicture}
\caption{The hierarchy of agents in a smart environment}
\label{fig:agent-model}
\end{figure}
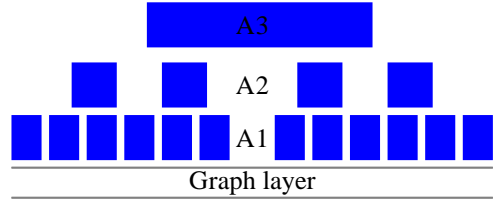

Let us consider a sample car park, c.f.\ Fig.~\ref{fig:graph-parking-space}.
It consists of some entrance/exit gates, and a number of identified parking spaces.
To achieve the goal mentioned above,
i.e.\ on-the-fly preference models,
we suggest to implement the following multi-agent system.
The world of things/objects is modeled using a graph structure,
that glues the cooperation of three types of agents into one
system. Fig.~\ref{fig:agent-model} shows the whole agent world,
i.e.\ agents that operate in the smart environment.
Existence of the following types of agents is assumed:
\begin{enumerate}
\item $A3$ -- agents also called \emph{decision agents},
      that permanently exist in the system and whose primary aim is to
      prepare/compute preference-based decisions for
      a new user/inhabitant entering the car park;
      these decisions are based on the gathered knowledge expressed
      in terms of logical specifications which are prepared by agents $A2$,
      decision agents can also modify knowledge, which is their secondary aim, when they find that
      the newly observed behaviors include contradictions in regard to the old behavior,
      i.e.\ knowledge expressed in (old) logical formulas,
      and that elimination of the contradiction might be a result of the formal analysis of logic formulas
      using, for example, the semantic tableaux method.
\item $A2$ -- agents also called \emph{follower agents},
      that might temporarily exist  in the system and whose aim is to
      observe objects that appear in the smart environments and
      build logical specifications considered
      as a set of temporal logic formulas that express behaviors of newly observed users/inhabitants;
      the logical specification constitutes knowledge about
      user preferences and is built basing on information form agents $A1$.
      They are generated when some event occurs.
\item $A1$ -- agents also called \emph{reactive agents}, or \emph{node agents},
      that exist permanently in the system and whose aim is to
      operate in an individual node,  gathering information about users/inhabitants
      who reach this node in the IoT network;
      information is obtained through sensors and combined with
      the identification (generally: RFID, PDA devices, biometric data, image scanning and pattern recognition, and others) of a user/inhabitant.
\end{enumerate}

The graph layer is defined as a labelled and attributed graph (abbrev.\ $LA$-graph) defined below.
\begin{definition}
\label{An-LA-graph-is}
An \emph{LA-graph} is a labelled and attributed
digraph of the following form $G=(V,E,\{lab_{X},att_{X}\}_{X=V,E})$, such that:
\begin{itemize}
  \item $V$ is a finite and nonempty set of vertices;
  \item $E\subset V\times V$ is a set of directed edges (arcs);
  \item $lab_{X}:X\rightarrow\mathcal{L}_{X}$ are labeling functions for
nodes ($X=V)$ and edges ($X=E)$ respectively, where $\mathcal{L}_{V},\mathcal{L}_{E}$
are sets of node and edge labels;
  \item $att{}_{X}:X\rightarrow2^{\mathcal{A}_{X}}$ are attributing functions
for nodes ($X=V)$ and edges ($X=E)$ respectively, where $\mathcal{A}_{V},\mathcal{A}_{E}$
are sets of node and edge attributes.
\end{itemize}
\end{definition}
The interpretation of labels and attributes in Definition~\ref{An-LA-graph-is}
is as follows. A label $l\in\mathcal{L}$ unambiguously identifies
a given vertex/edge, e.g. by assigning an unique name to an object;
an attribute $a\in\mathcal{A}$ is some property of a vertex/edge.
As stated in Definition~\ref{An-LA-graph-is}, one
may assign a set of attributes to a given entity. It should be stressed
that an attribute $a$ must not be confused with its value. Thus the
notion of $LA$-graph may be compared to a class definition. The graph
analog of a class instance is an instantiated LA-graph, defined
below.
\begin{definition}
Let $G=(V,E,\{lab_{X},att_{X}\}_{X=V,E})$ be an LA-graph.
An \emph{instantiation} of G is a triple $\hat{G}=(G,val_{V},val_{E})$, where $val_{X}:X\times\mathcal{A}_{X}\rightarrow\Omega_{X}$
is an instantiating function for nodes ($X=V$) and edges ($X=E$)
respectively.
$\hat{G}$ will be also referred to as an \emph{instantiated LA-graph} (shortly, ILA-graph).
\end{definition}
The mentioned idea will be explained using the example of the parking system.
The graph consists of only four types of nodes:
\begin {itemize}
\item node labelled by G -- that describe a gateway to the parking,
\item node labelled by R -- that describe an road segment,
\item node labelled by P -- that describe a parking space,
\item node labelled by C -- that describe a car.
\end {itemize}

In real solutions, we have to also consider a few types of sensors and the area of their cooperation,
but it will only influence  more complex behavior of the $A1$-type agent (so we will not consider them here).
We assume that  each node labelled by $G$, $R$ or $P$ has associated $A1$-type agents that discover
the appearance of a car in the space which it describes.
A more complex action is associated with the event of a car appearing in
the gateway (coming for outside of the parking); it consist of the sequence of actions:
\begin{itemize}
    \item a new node labelled $C$ is added to a graph -- it is linked with node labeled by $G$,
    \item a new agent of type $A2$ is created, and it communicates with the agent of type $A3$
    supervising this gateway -- asking for the preference of the identified car.
    This generates agents which follows the car, observing the driver behavior
    both while it travels to the parking space and while it leaves the parking.
    \item when the car leaves the car park, agent of the $A2$ type sends the observations to
    an agent of type $A3$, and destroys itself.
\end{itemize}

Let us consider a simple yet illustrative example for the approach.
Let us present rules for the $A1$ agents,
which are assigned to particular nodes of the parking space/graph structure.
The basic events that refer to the presence of users/inhabitants are recorded in nodes.
Let $O=\{ o_{1},o_{2},... \}$ is a set of users/inhabitants identified in the system.
Individual users have unique identifiers.
Let $D= \{ d_{1}, d_{2}, ... \}$ is a set of events,
where every $d_{i}$ belongs to $\langle O,V,T \rangle$,
where $O$ is a set of identified users/inhabitants,
$V$ is a node of a network,
and $T$ is a set of time stamps.
For example,
$d_{i}=\langle idOla91,p0018,t2014.01.28.09.30.15 \rangle$
means that the presence of the $idOla91$ object is observed at
the physical point/area $p0018$ of the parking space,
and the timestamp assigned to this event is $t2014.01.28.09.30.15$.

Let us present rules for the $A2$ agents,
which occupy the middle level in the entire hierarchy of the agent activities.
Agents gather knowledge about preferences of users/inhabitants in the considered area.
Preferences are expressed in terms of temporal logic formulas.
To obtain such logical specifications, the information produced by agents $A1$,
i.e.\ events registered in particular nodes are processed.
The $A2$ agents translate physical events to logical specifications.
The main idea for this is to analyze timestamps of events;
however, the detailed algorithm will be the goal of separate work,
and here only  brief information is presented.
The input for this translation are events $d_{i}$ as defined above.
The output are logical formulas understood as triples of the form
$l_{i}=\langle id,f,r \rangle$,
where $id$ is an identifier an object that operates in the parking space/IoT,
$f$ is a temporal logic formula, and
$r$ is the number of occurrences of this formula as a result of a user behavior.
The entire logical specification is a set of these triples,
i.e.\ $S=\{ l_{i} : i \geq 0 \}$.
The introduced notion requires some explanation.
The system stores information about different users and
the $id$ allows to differentiate formulas
intended for a particular user.
The meaning of $f$ is obvious,
i.e.\ it is a syntactically-correct temporal logic formula.
The $r$ element, where $r>0$, is a kind of counter and
it means multiple occurrences of a given formula as a result
of multiple observations of the same behavior in the past.
For example,
$\langle idOla91,g2 \imp\som p018,7 \rangle$ and
$\langle idOla91,g2 \imp\som p015,2 \rangle$
means that user $idOla91$ enters gate $g2$ and
sometimes reaches the parking area $p018$ (seven times in the past), and
sometimes reaches the parking area $p015$ (two times in the past).
When the preference decision is taken,
and if $p018$ is free,
then this parking area is suggested as the most preferred one,
otherwise $p015$ is suggested or, if it is not free, no suggestion is made.

Let us present rules for the $A3$ agents,
which occupy the highest level in the  hierarchy of the agent activities,
and whose purpose is to prepare preference decisions for a user/inhabitant.
Agents analyze knowledge about preferences expressed in terms of logical formulas,
which are produced by agents $A2$.
The input for this analysis are logical specification.
The output are preference decisions prepared for a particular user/inhabitant.

\begin{figure}[htb]
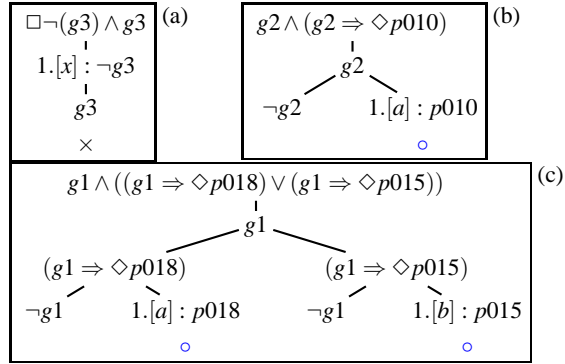

{\small
\begin{tabular}{ll}
\framebox{
\pstree[levelsep=4.0ex,nodesep=2pt,treesep=25pt]
       {\TR{$\alw\neg (g3) \con g3$}}
       {\pstree{\TR{$1.[x]: \neg g3$}}
       {\pstree{\TR{$g3$}}{$\times$}}}} (a)
\qquad
\framebox{
\pstree[levelsep=4.0ex,nodesep=2pt,treesep=25pt]
       {\TR{$g2 \con (g2 \imp \som p010)$}}
       {\pstree{\TR{$g2$}}
         {{\TR{$\neg g2$}}{\pstree{\TR{$1.[a]: p010$}}{\textcolor{blue}{$\circ$}}}}}} (b)\\
\framebox{
\pstree[levelsep=4.0ex,nodesep=2pt,treesep=25pt]
       {\TR{$g1 \con ((g1 \imp\som p018) \dis (g1 \imp\som p015))$}}
       {\pstree{\TR{$g1$}}
         {{\pstree{\TR{$(g1 \imp\som p018)$}}{{\TR{$\neg g1$}}{\pstree{\TR{$1.[a]:p018$}}{\textcolor{blue}{$\circ$}}}}}
         {\pstree{\TR{$(g1 \imp\som p015)$}}{{\TR{$\neg g1$}}{\pstree{\TR{$1.[b]:p015$}}{\textcolor{blue}{$\circ$}}}}}}}}
         (c)
\end{tabular}
}
\caption{The sample truth trees}
\label{fig:truth-trees}
\vspace{-.8\baselineskip}
\end{figure}
Let us consider some cases to explain the presented ideas.
Assume that the logical specification for user $o_{i}$ contains
a logical formula $\alw\neg(g3)$, which means that the user never entered gate $g3$.
However,
when at a certain time point user $o_{i}$ appears at $g3$,
then it provides the logical formula $\alw\neg (g3) \con g3$ which might give
the reasoning tree for the semantic tableaux method shown in
Fig.~\ref{fig:truth-trees}.a, c.f.\ closed branch ($\times$), i.e.\ a contradiction.
Of course,
this tree could be a part of a larger truth tree,
which is omitted here to simplify considerations.
It follows that the logical specification should be modified by
removing formula $\alw\neg (g3)$ from the initial specification,
then a new formula which results from a new event,
entering gate $g3$, is to be added to the specification.
Another case could refer to a situation when user enters gate $g2$
and the logical specification contains formula $g2 \imp\som p010$,
which means that if $g2$ is reached then sometime area $p010$ is reached.
It leads to the following formulas:
$g2 \con (g2 \imp\som p010) \Longrightarrow \som p010$,
or using the truth tree Fig.~\ref{fig:truth-trees}.b,
c.f.\ the open branch (\textcolor{blue}{$\circ$}).
The preference decision is the sample $p010$ parking space, if it is free.
The last case is the situation when a gate is reached
and there exist two (or more) different (sub-)formulas,
i.e.\ $g1 \con ((g1 \imp\som p018) \dis (g1 \imp\som p015)) \Longrightarrow \som p018 \dis \som p015$,
or using the truth tree Fig.~\ref{fig:truth-trees}.c,
c.f.\ the open branches (\textcolor{blue}{$\circ$}).
It means that both $p018$ and $p015$ are areas of preference.
It also means that the last element of the $l_{i}$ triple, which is frequency $r$ of
a particular formula determining which parking area is chosen as a preferred one,
if it is free,
i.e.\ the $r$ element does not influence the formal inference process but
it supports the choice between open branches
which are result of an inference.

\section{Conclusions}

In this work we present an example of application of the IoT concept in
a multi-agent environment where the external knowledge is represented by
a graph and the preference model is represented by formal logic.
In case of a real system, such a graph should be divided into smaller parts that will
cooperate with each other  (with an explicit synchronization mechanism).
Such an environment called Replicated Complementary graphs is supported by
the GRADIS~\cite{Kotulski-2008-iccs} multi-agent framework, where each agent controls one local graph $G_{i}$.
Following the FIPA~\cite{FIPA-2014} specification~\cite{FIPA-Specifications-2014}, we assume
a very simple functionality of a multi-agent environment,
reduced to a message transport and a broker system.
This approach is similar to those applied in popular frameworks like JADE~\cite{JADE-2014} or
Retsina~\cite{Retsina-2014}.

\bibliographystyle{ieeetr1}
\bibliography{rk-bib-rk,rk-bib-main,rk-bib-pervasive,rk-bib-graph}

\end{document}